\documentclass[aps,twocolumn]{revtex4}%
\usepackage{amsfonts}
\usepackage{amsmath}
\usepackage{amssymb}
\usepackage{graphicx}%
\providecommand{\U}[1]{\protect\rule{.1in}{.1in}}

\begin{document}
\title{Density matrix of the superposition of excitation on coherent states with
thermal light and its statistical properties }
\author{Li-yun Hu}
\affiliation{Department of Physics, Shanghai Jiao Tong University, Shanghai 200030, China}
\author{Hong-yi Fan}
\affiliation{Department of Physics, Shanghai Jiao Tong University, Shanghai 200030, China}

\begin{abstract}
A beam's density matrix that is described by the superposition of excitation
on coherent states with thermal noise (SECST) is presented, and its matrix
elements in Fock space are calculated. The maximum information transmitted by
the SECST beam is derived. It is more than that by coherent light beam and
increases as the excitation photon number increases. In addition, the
nonclassicality of density matrix is demonstrated by calculating its Wigner function.

\bigskip


\end{abstract}
\date{5 July 2008}
\maketitle

\section{Introduction}

Recently, much attention has been paid to the excitation on coherent
states (ECS) \cite{1,2,3,4,5,6}. As pointed out in Refs.\cite{2,3},
the single photon ECS causes a classical-to-quantum (nonclassical)\
transition. The ECSs can be considered as a generalization of
coherent states \cite{7,8} and number eigenstates. All these states
can be used as signal beams in optical communications field, in
which the nonclassicality of signals plays an important role.

However, in reality, signal beams are usually mixed with thermal noise.
Statistical properties of the superposition of (squeezed) coherent states with
thermal light (SCST) have been investigated by calculating the photon number
matrix elements $\left\langle N\right\vert \rho\left\vert M\right\rangle $ of
SCST's density matrix \cite{9,10}. These properties are useful in quantum
optics and quantum electronics (e.g. how lasers working well above threshold,
heterodyne detection of light, etc.) \cite{11}. Some general properties of the
density matrices which describe coherent, squeezed and number eigenstates in
thermal noise are studied in Ref.\cite{12}. It is found that the information
transmitted by the superposition of number eigenstates with thermal light
(SNET) beam is less than that by the SCST beam \cite{13}.

In this paper, we investigate statistical properties of the superposition of
ECS with thermal light (SECST). We present the relevant density matrix in Fock
space and derive the Mandel $Q$ parameter. The SECST field can exhibit a
significant amount of super-Poissonian photon statistics (PPS) due to the
presence of thermal noise for excitation photon number $m=0;$ while for
$m\neq0$ the SECST field can present the sub-PPS when the thermal mean photon
number is less than a threshold value. In addition, the threshold value
increases as $m$\ increases. We also calculate the maximum information
(channel capacity) transmitted by the SECST beam, which increases as $m$
increases. In addition, the nonclassicality of density matrix is also
presented by calculating the Wigner function of the SECST.

Our paper is arranged as follows. In Sec. II we present the density matrix
$\rho$ that describes the SECST and calculate its matrix elements in Fock
space by using the normal ordered form of $\rho$. The PPS distributions are
discussed in Sec III. The maximum information is calculated in Sec. IV. Sec. V
is devoted to deriving the Wigner function of the SECST and discussing its
nonclassicality in details. Conclusions are summarized in the last section.

\section{Excitation on coherent states with thermal noise}

Firstly, let us briefly review the excitation on coherent states (ECSs). The
ECSs, first introduced by Agarwal and Tara \cite{1}, are the result of
successive elementary one-photon excitations of a coherent state, and is an
intermediate state in between the Fock state and the coherent state, since it
exhibits the sub-Poissonian character. Theoretically, the ECSs can be obtained
by repeatedly operating the photon creation operator $a^{\dag}$\ on a coherent
state, so its density operator is {\small \ }
\begin{equation}
\rho_{0}=C_{\alpha,m}a^{\dag m}\left\vert \alpha\right\rangle \left\langle
\alpha\right\vert a^{m}, \label{1}%
\end{equation}
where $C_{\alpha,m}=[m!L_{m}(-\left\vert \alpha\right\vert ^{2})]^{-1}$ is the
normalization factor, $\left\vert \alpha\right\rangle =\exp(-\left\vert
\alpha\right\vert ^{2}/2+\alpha a^{\dagger})\left\vert 0\right\rangle $ is the
coherent state \cite{7,8}, and $L_{m}\left(  x\right)  $ is the $m$th-order
Laguerre polynomial.

The SECST is described by the density matrix \cite{12}%
\begin{align}
\rho &  =\int\frac{d^{2}z}{\pi}P\left(  z\right)  D\left(  z\right)  \rho
_{0}D^{\dag}\left(  z\right)  ,\label{2}\\
P\left(  z\right)   &  =\frac{1}{\bar{n}_{t}}\exp\left[  -\frac{\left\vert
z\right\vert ^{2}}{\bar{n}_{t}}\right]  ,\label{3}%
\end{align}
where $D\left(  z\right)  =\exp(za^{\dag}-z^{\ast}a)$ is the displacement
operator, and $\bar{n}_{t}$ is the mean number of thermal photons for
$\rho_{0}\rightarrow\left\vert 0\right\rangle \left\langle 0\right\vert $. We
can easily prove that $\mathtt{Tr}\rho=1,$ as it should be. In fact,
\begin{align}
\mathtt{Tr}\rho &  =\int\frac{d^{2}z}{\pi}P\left(  z\right)  \mathtt{Tr}%
\left[  D\left(  z\right)  \rho_{0}D^{\dag}\left(  z\right)  \right]
\nonumber\\
&  =\int\frac{d^{2}z}{\pi}P\left(  z\right)  \mathtt{Tr}\left(  \rho
_{0}\right)  \nonumber\\
&  =\int\frac{d^{2}z}{\pi}P\left(  z\right)  =1.\label{4}%
\end{align}

\subsection{Normal ordering form of the SECST}

For the simplicity in our later calculation, we first perform the integration
in Eq.(\ref{2}) by using the technique of integration within an ordered
product (IWOP) of operators \cite{14,15}. Using the normal ordering form of
the vacuum projector $\left\vert 0\right\rangle \left\langle 0\right\vert
=\colon\exp(-a^{\dag}a)\colon,$ we can reform Eq.(\ref{2}) as the following
form%
\begin{align}
\rho &  =C_{\alpha,m}e^{-\left\vert \alpha\right\vert ^{2}}\int\frac{d^{2}%
z}{\pi}P\left(  z\right)  D\left(  z\right)  \colon a^{\dag m}\nonumber\\
&  \times\exp\left(  \alpha a^{\dag}+\alpha^{\ast}a-a^{\dag}a\right)
a^{m}\colon D^{\dag}\left(  z\right)  \nonumber\\
&  =\frac{C_{\alpha,m}}{\bar{n}_{t}}\colon\exp\left(  -\left\vert
\alpha\right\vert ^{2}-a^{\dag}a+\allowbreak a^{\dag}\alpha+a\alpha^{\ast
}\right)  \nonumber\\
&  \times\int\frac{d^{2}z}{\pi}\exp\left[  -\frac{1+\bar{n}_{t}}{\bar{n}_{t}%
}\left\vert z\right\vert ^{2}\right]  \nonumber\\
&  \times\exp\left[  \left(  a^{\dag}-\alpha^{\ast}\right)  \allowbreak
z+\left(  a-\alpha\right)  z^{\ast}\right]  \nonumber\\
&  \times\left(  a^{\dag}-z^{\ast}\right)  ^{m}\left(  a-z\right)  ^{m}%
\colon.\label{5}%
\end{align}
In the last step of (\ref{5}), we noticed that for any operator $f(a^{\dag
},a)$%
\begin{equation}
D\left(  z\right)  f(a^{\dag},a)D^{\dag}\left(  z\right)  =f(a^{\dag}-z^{\ast
},a-z).\label{6}%
\end{equation}
Making two independent variable displacements,%
\[
a^{\dag}-z^{\ast}\rightarrow\beta^{\ast},a-z\rightarrow\beta,
\]
(note that operators $a^{\dag},a$ can be considered as C-number within the
normal order $\colon\colon$), thus Eq.(\ref{5}) can be rewritten as%
\begin{align}
\rho &  =\frac{C_{\alpha,m}}{\bar{n}_{t}}\colon\exp\left(  -\left\vert
\alpha\right\vert ^{2}-\frac{1}{\bar{n}_{t}}a^{\dag}a\right)  \nonumber\\
&  \times\int\frac{d^{2}\beta}{\pi}\beta^{\ast m}\beta^{m}\exp\left[
-\lambda_{t}^{-2}\left\vert \beta\right\vert ^{2}\right.  \nonumber\\
&  \left.  +\left(  \allowbreak\alpha^{\ast}+\allowbreak\frac{a^{\dag}}%
{\bar{n}_{t}}\right)  \beta+\left(  \frac{a}{\bar{n}_{t}}+\alpha\right)
\beta^{\ast}\right]  \colon\nonumber\\
&  =\frac{C_{\alpha,m}}{\bar{n}_{t}}\lambda_{t}^{2m+2}\colon\exp\left(
-\left\vert \alpha\right\vert ^{2}-\frac{1}{\bar{n}_{t}}a^{\dag}a\right)
\nonumber\\
&  \times\int\frac{d^{2}\beta}{\pi}\beta^{\ast m}\beta^{m}\exp\left[
-\left\vert \beta\right\vert ^{2}+A^{\dag}\beta+A\beta^{\ast}\right]
\colon,\label{7}%
\end{align}
where we have set $\lambda_{t}=\sqrt{\bar{n}_{t}/(1+\bar{n}_{t})}$ and
$A=\lambda_{t}(\frac{1}{\bar{n}_{t}}a+\alpha).$ Then using the integration
expression of two-variable Hermite polynomial $H_{m,n}$ \cite{16},
\begin{align}
&  (-1)^{n}e^{-\xi\eta}H_{m,n}\left(  \xi,\eta\right)  \nonumber\\
&  =\int\frac{d^{2}z}{\pi}z^{n}z^{\ast m}\exp\left[  -\left\vert z\right\vert
^{2}+\xi z-\eta z^{\ast}\right]  ,\label{8}%
\end{align}
we can put Eq.(\ref{7}) into
\begin{align}
\rho &  =\frac{C_{\alpha,m}}{\bar{n}_{t}}\lambda_{t}^{2m+2}\colon\left(
-1\right)  ^{m}H_{m,m}\left(  A^{\dag},-A\right)  \nonumber\\
&  \times\exp\left[  -\frac{\left(  a-\alpha\right)  \left(  a^{\dag}%
-\alpha^{\ast}\right)  }{\bar{n}_{t}+1}\right]  \colon.\label{9}%
\end{align}
In particular, when $m=0$, corresponding to the case of superposition of
coherent state with thermal noise, Eq.(\ref{9}) reduces to
\begin{equation}
\rho=\frac{1}{\bar{n}_{t}+1}D\left(  \alpha\right)  \colon e^{-\frac{a^{\dag
}a}{\bar{n}_{t}+1}}\colon D^{\dag}\left(  \alpha\right)  ,\label{9.1}%
\end{equation}
which can be directly checked by using Eqs.(\ref{2}) and (\ref{3}) as well as
noticing $\rho_{0}=\left\vert \alpha\right\rangle \left\langle \alpha
\right\vert .$

Further employing the relation between Hermite polynomial and Laguerre
polynomial \cite{16},%
\begin{equation}
H_{m,n}\left(  \xi,\kappa\right)  =\left\{
\begin{array}
[c]{cc}%
n!\left(  -1\right)  ^{n}\xi^{m-n}L_{n}^{m-n}\left(  \xi\kappa\right)  , &
m>n\\
m!\left(  -1\right)  ^{m}\kappa^{n-m}L_{m}^{n-m}\left(  \xi\kappa\right)  , &
m<n
\end{array}
\right.  , \label{10}%
\end{equation}
we can see that
\begin{align}
\rho &  =\frac{1}{L_{m}(-\left\vert \alpha\right\vert ^{2})}\frac{\bar{n}%
_{t}^{m}}{(1+\bar{n}_{t})^{m+1}}\colon L_{m}\left(  -A^{\dag}A\right)
\nonumber\\
&  \times\exp\left[  -\frac{\left(  a-\alpha\right)  \left(  a^{\dag}%
-\alpha^{\ast}\right)  }{\bar{n}_{t}+1}\right]  \colon. \label{11}%
\end{align}
Eqs.(\ref{9}) and (\ref{11}) are the normal ordering form of the SECST. From
these it is convenient to calculate the phase space distributions, such as
Q-function, P-representation and Wigner function.

\subsection{The matrix elements $\left\langle N\right\vert \rho\left\vert
M\right\rangle $}

Now we calculate the matrix elements of $\rho$ in Eq.(\ref{2}) between two
number states $\left\langle N\right\vert $ and $\left\vert M\right\rangle ,$
i.e., $\left\langle N\right\vert \rho\left\vert M\right\rangle .$ Employing
the overcompleteness of coherent states, one can express the matrix elements
$\left\langle N\right\vert \rho\left\vert M\right\rangle $ as
\begin{equation}
\left\langle N\right\vert \rho\left\vert M\right\rangle =\int\frac{d^{2}\beta
d^{2}\gamma}{\pi^{2}}\left\langle N\right.  \left\vert \beta\right\rangle
\left\langle \beta\right\vert \rho\left\vert \gamma\right\rangle \left\langle
\gamma\right.  \left\vert M\right\rangle ,\label{12}%
\end{equation}
where the overlap between the coherent state and the number state is given by%
\begin{equation}
\left\langle \gamma\right.  \left\vert M\right\rangle =\frac{1}{\sqrt{M!}%
}e^{-\left\vert \gamma\right\vert ^{2}/2}\gamma^{\ast M},\label{13}%
\end{equation}
and the matrix elements $\left\langle \beta\right\vert \rho\left\vert
\gamma\right\rangle $ can be obtained from Eq.(\ref{9}) due to $\rho^{\prime}%
$s normal ordering form,%
\begin{align}
\left\langle \beta\right\vert \rho\left\vert \gamma\right\rangle  &  =\left(
-1\right)  ^{m}\frac{C_{\alpha,m}}{\bar{n}_{t}}\lambda_{t}^{2m+2}%
e^{-\left\vert \alpha\right\vert ^{2}/(\bar{n}_{t}+1)}\nonumber\\
&  \times\frac{\partial^{2m}}{\partial\tau^{m}\partial\tau^{\prime m}}%
\exp\left[  -\tau\tau^{\prime}+\lambda_{t}\tau\alpha^{\ast}-\lambda_{t}%
\alpha\tau^{\prime}\right]  \nonumber\\
&  \times\exp\left\{  \left(  \frac{\alpha+\bar{n}_{t}\allowbreak\gamma}%
{\bar{n}_{t}+1}+\frac{\lambda_{t}\tau}{\bar{n}_{t}}\right)  \beta^{\ast}%
-\frac{1}{2}\left\vert \beta\right\vert ^{2}\right.  \nonumber\\
&  -\left.  \frac{1}{2}\left\vert \gamma\right\vert ^{2}+\left(  \frac
{\alpha^{\ast}}{\bar{n}_{t}+1}-\frac{\lambda_{t}\tau^{\prime}}{\bar{n}_{t}%
}\right)  \gamma\right\}  _{\tau=\tau^{\prime}=0},\label{14}%
\end{align}
where we have used the generating function of two-variable Hermite polynomial
$H_{m,n},$%
\begin{equation}
H_{m,n}\left(  x,y\right)  =\left.  \frac{\partial^{m+n}}{\partial
t^{m}\partial t^{\prime n}}\exp\left[  -tt^{\prime}+tx+t^{\prime}y\right]
\right\vert _{t=t^{\prime}=0}.\label{15}%
\end{equation}
When $M=N,$ $\left\langle N\right\vert \rho\left\vert N\right\rangle $ is just
the photon number distribution of the SECST. Then combing with Eqs.(\ref{14}),
(\ref{12}) and (\ref{13}), after a lengthy but straightforward calculation,
one can get the matrix elements $\left\langle N\right\vert \rho\left\vert
M\right\rangle ,$ (without loss of generality, let $M\geqslant N$)%
\begin{align}
\left\langle N\right\vert \rho\left\vert M\right\rangle  &  =\frac{\left(
-1\right)  ^{N}}{\sqrt{M!N!}}\frac{\lambda_{t}^{2N}C_{\alpha,m}}{\bar{n}%
_{t}+1}e^{-\left\vert \alpha\right\vert ^{2}}\frac{\partial^{2m}}%
{\partial\upsilon^{m}\partial\upsilon^{\prime m}}\nonumber\\
&  .\left\{  e^{\lambda_{t}^{2}\upsilon\upsilon^{\prime}}H_{M,N}\left(
\frac{\upsilon^{\prime}}{\bar{n}_{t}+1},-\frac{\upsilon}{\bar{n}_{t}}\right)
\right\}  _{\upsilon=\alpha,\upsilon^{\prime}=\alpha^{\ast}},\label{16}%
\end{align}
where we have used the integral formula \cite{17}
\begin{equation}
\int\frac{d^{2}\beta}{\pi}f\left(  \beta^{\ast}\right)  \exp\left\{
-\left\vert \beta\right\vert ^{2}+\tau\beta\right\}  =f\left(  \tau\right)
,\label{17}%
\end{equation}
and another expression of two-variable Hermite polynomial $H_{m,n},$%
\begin{equation}
H_{m,n}\left(  \xi,\kappa\right)  =\sum_{l=0}^{\min(m,n)}\frac{m!n!\left(
-1\right)  ^{l}\xi^{m-l}\kappa^{n-l}}{l!\left(  n-l\right)  !\left(
m-l\right)  !}.\label{18}%
\end{equation}
In particular, when $m=0$, noticing $M\geqslant N$ and Eq.(\ref{10}),
Eq.(\ref{16}) reduces to%
\begin{align}
\left\langle N\right\vert \rho\left\vert M\right\rangle  &  =\sqrt{\frac
{N!}{M!}}\alpha^{\ast M-N}\frac{\left(  \bar{n}_{t}\right)  ^{N}}{\left(
\bar{n}_{t}+1\right)  ^{M+1}}\nonumber\\
&  \times e^{-\left\vert \alpha\right\vert ^{2}/(\bar{n}_{t}+1)}L_{N}%
^{M-N}\left[  -\frac{\left\vert \alpha\right\vert ^{2}}{\bar{n}_{t}\left(
\bar{n}_{t}+1\right)  }\right]  ,\label{19}%
\end{align}
which is just the Glauber-Lachs formula \cite{9} when $\bar{n}_{t}%
=(e^{\beta\omega}-1)^{-1}$. While for $\alpha=0,$ corresponding to the case of
superposition of number state with thermal light, using Eq.(\ref{18}),
Eq.(\ref{16}) becomes%
\begin{equation}
\left\langle N\right\vert \rho\left\vert M\right\rangle =\delta_{M,N}%
P_{N},\label{e20}%
\end{equation}
where ($k_{0}=\max[0,m-N]$)
\begin{equation}
P_{N}=\frac{m!N!}{\bar{n}_{t}+1}\sum_{k=k_{0}}^{m}\frac{1}{k!}\frac{\left(
\frac{\bar{n}_{t}}{\bar{n}_{t}+1}\right)  ^{k+N}\left[  \bar{n}_{t}\left(
\bar{n}_{t}+1\right)  \right]  ^{k-m}}{\left(  k+N-m\right)  !\left[
(m-k)!\right]  ^{2}}.\label{e21}%
\end{equation}
Eq.(\ref{e20}) is just the result of Ref. \cite{13}.

\section{Sub-Poissonian photon statistics}

To see clearly the photon statistics properties of the SECST, in this section,
we pay our attention to the variance of the photon number operator
$\left\langle \left(  \Delta\hat{n}\right)  ^{2}\right\rangle =\left\langle
\hat{n}^{2}\right\rangle -\left\langle \hat{n}\right\rangle ^{2}.$ In
particular, we will examine the evolution of the Mandel $Q$ parameter defined
as%
\begin{align}
Q  &  =\frac{\left\langle \left(  a^{\dag}a\right)  ^{2}\right\rangle
}{\left\langle a^{\dag}a\right\rangle }-\left\langle a^{\dag}a\right\rangle
\nonumber\\
&  =\frac{\left\langle a^{2}a^{\dag2}\right\rangle -\left\langle aa^{\dag
}\right\rangle ^{2}-\left\langle aa^{\dag}\right\rangle }{\left\langle
aa^{\dag}\right\rangle -1}, \label{26}%
\end{align}
which measures the derivation of the variance of the photon number
distribution of the field state under consideration from the
Poissonian distribution of the coherent state. $Q=1,Q>1$ and $Q<1$
correspond to Poissonian photon statistics (PPS), super-PPS and
sub-PPS, respectively.

In order to calculate the average value in Eq.(\ref{26}), we first calculate
the value of $\left\langle \alpha\right\vert a^{n}a^{\dag m}\left\vert
\alpha\right\rangle .$ In fact, using
\begin{equation}
\left\langle \alpha\right\vert a^{m+n}a^{\dag m+n}\left\vert \alpha
\right\rangle =\left(  m+n\right)  !L_{m+n}(-\left\vert \alpha\right\vert
^{2}) \label{27}%
\end{equation}
and
\begin{equation}
\int\frac{d^{2}z}{\pi}z^{n}z^{\ast m}P\left(  z\right)  =\bar{n}_{t}%
^{m}m!\delta_{m,n}, \label{28}%
\end{equation}
we can evaluate (for writing's convenience, let $L_{m}$ denote $L_{m}%
(-\left\vert \alpha\right\vert ^{2})$)
\begin{equation}
\left\langle a^{\dag}a\right\rangle =\frac{1+m}{L_{m}}L_{m+1}+\bar{n}_{t}-1,
\label{29}%
\end{equation}
and%
\begin{equation}
\left\langle a^{2}a^{\dag2}\right\rangle =2\bar{n}_{t}^{2}+\frac{m+1}{L_{m}%
}\left[  4\bar{n}_{t}L_{m+1}+\left(  m+2\right)  L_{m+2}\right]  . \label{30}%
\end{equation}
Substituting Eqs.(\ref{29}) and (\ref{30}) into (\ref{26}) leads to%
\begin{align}
Q  &  =\frac{\bar{n}_{t}\left(  \bar{n}_{t}-1\right)  L_{m}+\left(  2\bar
{n}_{t}-1\right)  \left(  m+1\right)  L_{m+1}}{\left(  1+m\right)
L_{m+1}+\left(  \bar{n}_{t}-1\right)  L_{m}}\nonumber\\
&  +\frac{(m+1)(m+2)L_{m+2}-\frac{\left(  m+1\right)  ^{2}}{L_{m}}L_{m+1}^{2}%
}{\left(  1+m\right)  L_{m+1}+\left(  \bar{n}_{t}-1\right)  L_{m}}. \label{31}%
\end{align}
At the zero-temperature limit $(\bar{n}_{t}\rightarrow0)$, Eq.(\ref{31}) just
reduces to Eq.(2.20) in Ref.\cite{1}. \begin{figure}[ptb]
\label{Fig1} \centering\includegraphics[width=7cm]{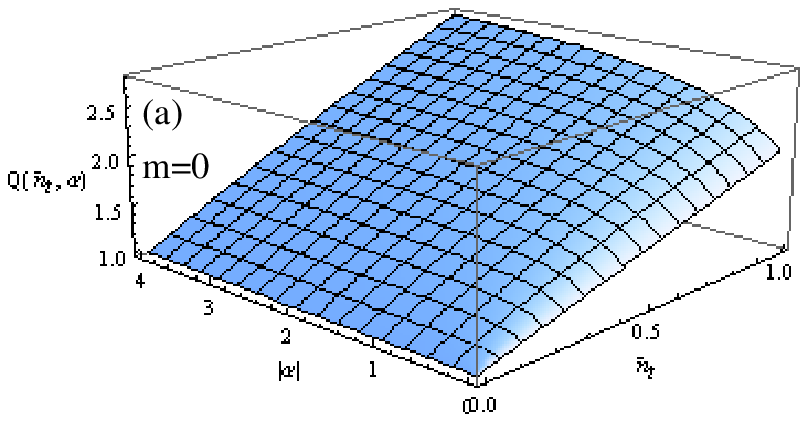} \label{Fig2}%
\centering\includegraphics[width=7cm]{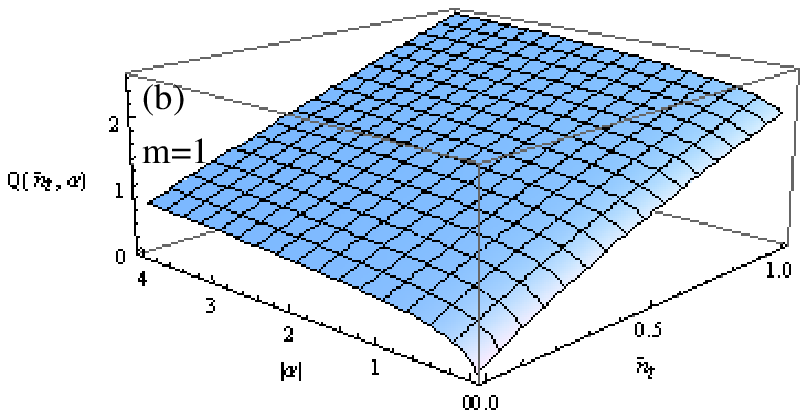} \label{Fig3}%
\centering\includegraphics[width=7cm]{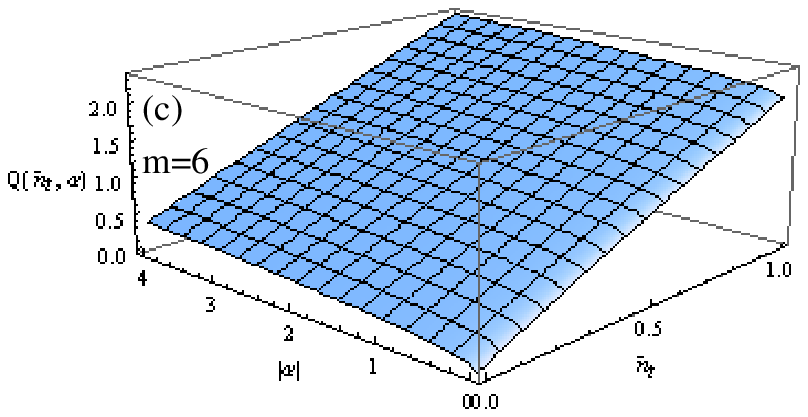}\caption{{\protect\small (Color
online)} {\protect\small The evolution of Mandel $Q$ parameter as a function
of (}$n_{t},\left\vert \alpha\right\vert ${\protect\small ) for different
values }$m.$}%
\end{figure}

In Fig.1, we display the parameter $Q\left(  n_{t},\left\vert \alpha
\right\vert \right)  $ as a function of $\left(  n_{t},\left\vert
\alpha\right\vert \right)  $ for different values $m.$ From Fig.1, we see
that, for the excitation photon number $m=0$ (see Fig.1 (a)), $Q\left(
\bar{n}_{t}=0,\left\vert \alpha\right\vert \right)  =1$ corresponding to
coherent state (a PPS); while $Q\left(  \bar{n}_{t}\neq0,\left\vert
\alpha\right\vert \right)  >1$, i.e., the SECST field exhibits a significant
amount of super-PPS due to the presence of $\bar{n}_{t}$. From Fig.1 (b) and
(c), we see that, when $m\neq0,$ the SECST field presents the sub-PPS when
$\bar{n}_{t}$ is less than a threshold value for a given $\left\vert
\alpha\right\vert ;$ the threshold value increases as $m$\ increases. For
example, when $\left\vert \alpha\right\vert =0,$ the threshold values are
about 0.414 and 0.481, respectively, for $m=1$ and $m=6$.

\section{Information transmitted by the SECST beam}

According to the negentropy principle of Brillouin \cite{18}, the maximum
information $I$ transmitted by a beam is
\begin{equation}
I=S_{\max}-S_{act}, \label{e32}%
\end{equation}
in which $S_{\max}$ and $S_{act}$ represent the maximum entropy and the actual
entropy, respectively, possessed by the quantum mechanical system described by
a density matrix $\rho$. Here the maximum information $I$ is an ideal one
transmitted through an ideal optical communication system.

For the SECST system, the actual entropy is
\begin{equation}
S_{act}=-\mathtt{Tr}\left(  \rho\ln\rho\right)  =-\sum_{N}\sigma_{N}\ln
\sigma_{N},\label{e33}%
\end{equation}
where $\rho=\sum_{N}\sigma_{N}\left\vert N\right\rangle \left\langle
N\right\vert ,$ and $\sigma_{N}=\left\langle N\right\vert \rho\left\vert
N\right\rangle .$ $\sigma_{N}$ can be obtained from Eq.(\ref{16}), i.e.,%
\begin{align}
\sigma_{N} &  =\frac{\bar{n}_{t}^{N}e^{-\left\vert \alpha\right\vert ^{2}%
}C_{\alpha,m}}{\left(  \bar{n}_{t}+1\right)  ^{N+1}}\frac{\partial^{2m}%
}{\partial\upsilon^{m}\partial\upsilon^{\prime m}}\nonumber\\
&  \times\left\{  e^{\lambda_{t}^{2}\upsilon\upsilon^{\prime}}L_{N}\left(
\frac{-\upsilon\upsilon^{\prime}}{\bar{n}_{t}\left(  \bar{n}_{t}+1\right)
}\right)  \right\}  _{\upsilon=\alpha,\upsilon^{\prime}=\alpha^{\ast}%
},\label{e34}%
\end{align}
which is independent of the phase of $\alpha.$ On the other hand, for a system
in thermal equilibrium, described by the density matrix $\rho_{th}$, with mean
photons number $\bar{n}_{t}$, its entropy is%
\begin{equation}
S=-\sum_{N}P_{N}\ln P_{N}=\ln(1+\bar{n}_{t})+\bar{n}_{t}\ln\frac{\bar{n}%
_{t}+1}{\bar{n}_{t}},\label{e35}%
\end{equation}
where $P_{N}=\bar{n}_{t}^{N}/\left(  \bar{n}_{t}+1\right)  ^{N=1}$ obtained
from Eq.(\ref{19}) under the condition $m=0,\alpha=0.$ Note that the maximum
entropy of the system is equal to the entropy of a system in thermal
equilibrium, with an equal mean number of photons. The mean photons number of
the SECST is given by Eq.(\ref{29}). Therefore, using Eq.(\ref{e35}), we have
\begin{align}
S_{\max} &  =\ln\left(  \left(  1+m\right)  \frac{L_{m+1}}{L_{m}}+\bar{n}%
_{t}\right)  \nonumber\\
&  +\left(  \left(  1+m\right)  \frac{L_{m+1}}{L_{m}}+\bar{n}_{t}-1\right)
\nonumber\\
&  \times\ln\left(  \frac{\left(  1+m\right)  L_{m+1}+\bar{n}_{t}L_{m}%
}{\left(  1+m\right)  L_{m+1}+\left(  \bar{n}_{t}-1\right)  L_{m}}\right)
.\label{e36}%
\end{align}

\begin{figure}[ptb]
\label{Fig7} \centering\includegraphics[width=7cm]{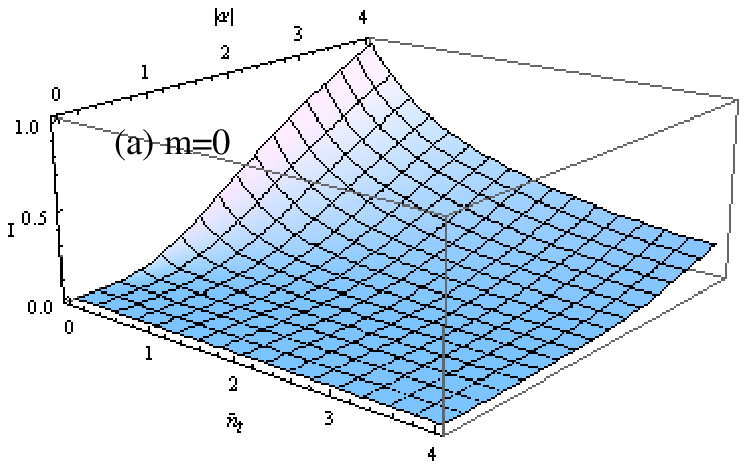} \label{Fig8}%
\centering\includegraphics[width=7cm]{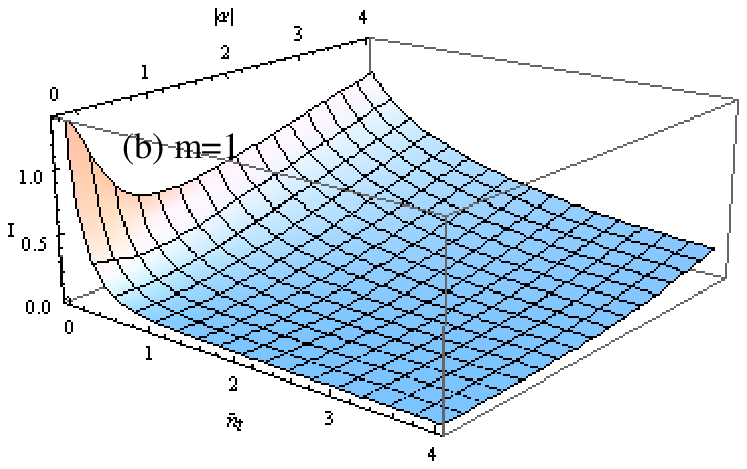}\caption{{\protect\small (Color
online)} {\protect\small The maximum information }$I\left(  \bar{n}%
_{t},\left\vert \alpha\right\vert \right)  ${\protect\small as a function of
}$\left(  \bar{n}_{t},\left\vert \alpha\right\vert \right)  $%
{\protect\small \ for some different values (a) }$m=0,(b)$ $m=1,$%
{\protect\small (truncating the infinite sum at }$N_{\max}=70$%
{\protect\small ).}}%
\end{figure}\begin{figure}[ptb]
\label{Fig9}
\centering\includegraphics[width=7cm]{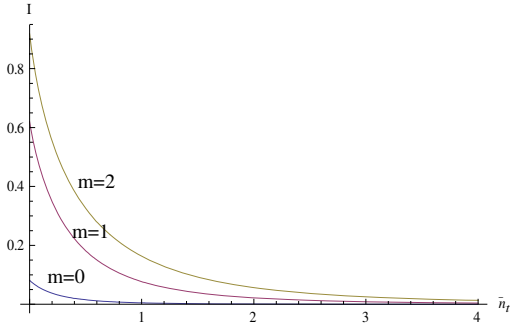}\caption{{\protect\small (Color
online) The maximum information }$I\left(  \bar{n}_{t},\left\vert
\alpha\right\vert =1\right)  ${\protect\small as a function of }$\left(
\bar{n}_{t}\right)  ${\protect\small \ for some different values (a)
}$m=0,(b)$ $m=1,(c)$ ${\protect\small m=2}$ {\protect\small (truncating the
infinite sum at }$N_{\max}=70${\protect\small ).}}%
\end{figure}

From Eqs.(\ref{e32}), (\ref{e33}) and (\ref{e36}), we can calculate the
maximum information transmitted by the SECST beam. In Fig. 2, the maximum
information $I\left(  \bar{n}_{t},\left\vert \alpha\right\vert \right)  $ is
plotted as a function of $\left(  \bar{n}_{t},\left\vert \alpha\right\vert
\right)  $\ for some different values $m$ (truncating the infinite sum at
$N_{\max}=70$). From Fig.2, we can see that, for a given $\bar{n}_{t},$
$I\left(  \bar{n}_{t},\left\vert \alpha\right\vert \right)  $ increases as the
value $\left\vert \alpha\right\vert $ increases; for given $\left\vert
\alpha\right\vert ,$ in general, $I\left(  \bar{n}_{t},\left\vert
\alpha\right\vert \right)  $ grows up as the value $\bar{n}_{t}$ increases. In
order to see clearly the effect of different parameter $m$ to $I\left(
\bar{n}_{t},\left\vert \alpha\right\vert \right)  ,$ we presented a plot in
Fig.3, from which it is obvious that $I\left(  \bar{n}_{t},\left\vert
\alpha\right\vert \right)  $ becomes bigger due to the presence of $m$, and
increases as $m$ increases. In other words, the maximum information
transmitted by the SECST beam is larger than that by the SCST ($m=0$). The
channel of ECS can carry with more information than that of coherent state. In
Ref. \cite{13}, Vourdas has pointed out that the coherent signals (of known
phase) can transmit more information than the number eigenvectors signals.
Thus among these three beams, the SECST beam can transmit most information.

\section{The Wigner function of the SECST}

\subsection{The Wigner function}

The Wigner function (WF) plays an important role in quantum optics, especially
the WF can be reconstructed from measurements \cite{19,20}. The WF is a
powerful tool to investigate the nonclassicality of optical fields. The
presence of negativity in the WF of optical field is a signature of its
nonclassicality is often used to describe the decoherence of quantum states.
In this section, using the normally ordered form of the SECST, we evaluate its
WF. For a single-mode system, the WF is given by \cite{21}%
\begin{equation}
W\left(  \gamma,\gamma^{\ast}\right)  =\frac{e^{2\left\vert \gamma\right\vert
^{2}}}{\pi}\int\frac{d^{2}\beta}{\pi}\left\langle -\beta\right\vert
\rho\left\vert \beta\right\rangle e^{2\left(  \beta^{\ast}\gamma-\beta
\gamma^{\ast}\right)  },\label{20}%
\end{equation}
where $\left\vert \beta\right\rangle $ is the coherent state and $\gamma
=x+iy$. From Eq.(\ref{20}) it is easy to see that once the normal ordered form
of $\rho$ is known, we can conveniently obtain the WF of $\rho.$

On substituting Eq.(\ref{9}) into Eq.(\ref{20}) we obtain the WF of the SECST%
\begin{align}
W\left(  \gamma,\gamma^{\ast}\right)   &  =\frac{\left(  \lambda_{t}^{2}%
A_{1}^{2}\right)  ^{m}C_{\alpha,m}}{\pi\left(  2\bar{n}_{t}+1\right)  }%
\exp\left\{  -\frac{2\left\vert \alpha-\gamma\right\vert ^{2}}{2n_{t}%
+1}\right\} \nonumber\\
&  \times\left(  -1\right)  ^{m}H_{m,m}\left(  \frac{A_{2}^{\ast}}{A_{1}%
},-\frac{A_{2}}{A_{1}}\right) \nonumber\\
&  =\frac{\left(  \lambda_{t}^{2}A_{1}^{2}\right)  ^{m}\exp\left\{
-\frac{2\left\vert \alpha-\gamma\right\vert ^{2}}{2\bar{n}_{t}+1}\right\}
}{\pi\left(  2\bar{n}_{t}+1\right)  L_{m}\left(  -\left\vert \alpha\right\vert
^{2}\right)  }L_{m}\left(  -\left\vert A_{2}\right\vert ^{2}/A_{1}^{2}\right)
, \label{21}%
\end{align}
where we have set%
\begin{align}
A_{1}^{2}  &  =1-\frac{1}{\left(  2\bar{n}_{t}+1\right)  \bar{n}_{t}%
},\nonumber\\
A_{2}  &  =\frac{\lambda_{t}\left(  \bar{n}_{t}+1\right)  }{\left(  2\bar
{n}_{t}+1\right)  \bar{n}_{t}}\left(  2\bar{n}_{t}\alpha-\alpha+2\gamma
\right)  . \label{22}%
\end{align}

\begin{figure}[ptb]
\label{Fig4} \centering\includegraphics[width=7cm]{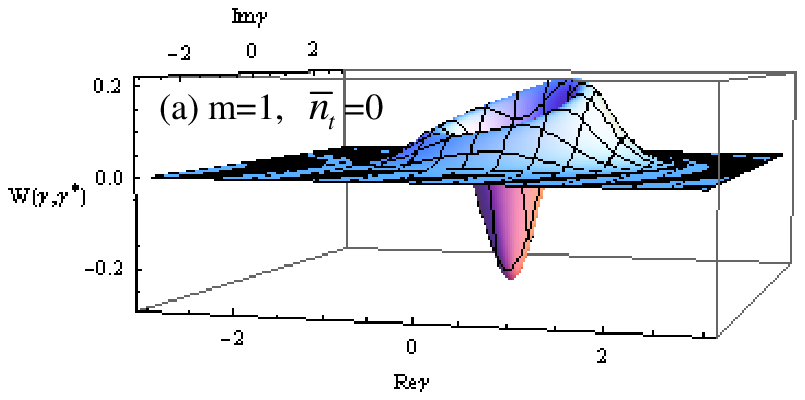} \label{Fig5}%
\centering\includegraphics[width=7cm]{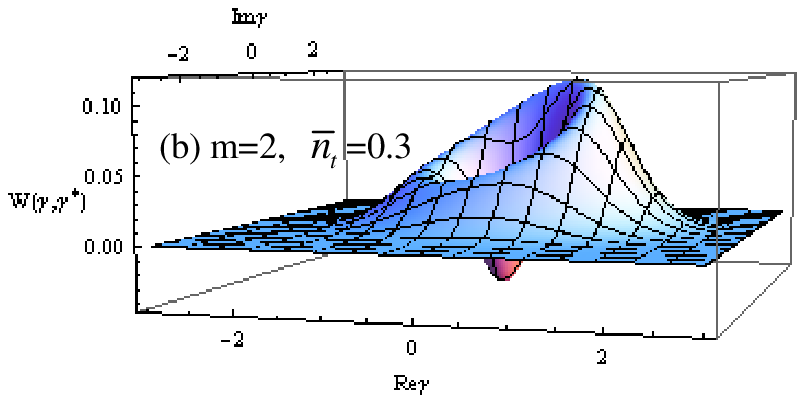} \label{Fig6}%
\centering\includegraphics[width=7cm]{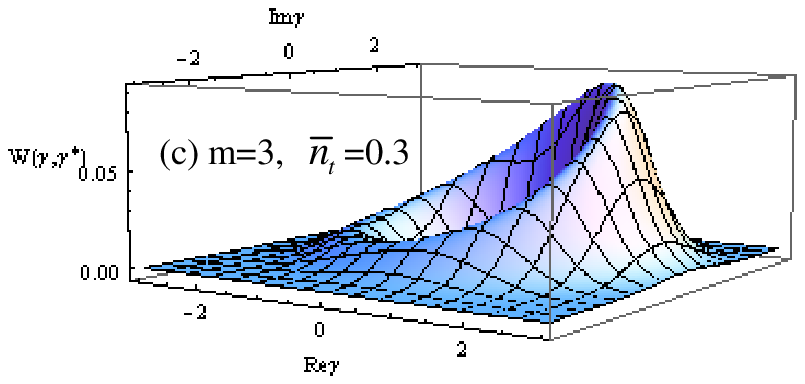}\caption{{\protect\small (Color
online)} {\protect\small The evolution of the Wigner function of the SECST
with }$\alpha=0.2+0.2i$ {\protect\small for several different values }$m$
{\protect\small and }$\bar{n}_{t}.$}%
\end{figure}Noticing that $L_{m}(-\left\vert \alpha\right\vert ^{2})>0,$ and
$L_{m}[-\left\vert A_{2}\right\vert ^{2}/A_{1}^{2}]>0$ when $1-\frac
{1}{\left(  2\bar{n}_{t}+1\right)  \bar{n}_{t}}>0,$ thus the WF of the SECST
is always positive under the condition of $\bar{n}_{t}>1/2$. In particular,
when $m=0$, \ Eq.(\ref{21}) becomes
\begin{equation}
W\left(  \gamma,\gamma^{\ast}\right)  =\frac{1}{\pi\left(  2\bar{n}%
_{t}+1\right)  }\exp\left\{  -\frac{2\left\vert \alpha-\gamma\right\vert ^{2}%
}{2\bar{n}_{t}+1}\right\}  ,\label{23}%
\end{equation}
which corresponds to the thermal state with mean photon number $\bar{n}_{t}$.
While for $\alpha=0,$ $A_{2}\rightarrow2\gamma\lambda_{t}\left(  \bar{n}%
_{t}+1\right)  /[\left(  2\bar{n}_{t}+1\right)  \bar{n}_{t}],$ $\left\vert
A_{2}\right\vert ^{2}/A_{1}^{2}\rightarrow4\left\vert \gamma\right\vert
^{2}\left(  \bar{n}_{t}+1\right)  /\{\left(  2\bar{n}_{t}+1\right)  \left[
\left(  2\bar{n}_{t}+1\right)  \bar{n}_{t}-1\right]  \}\equiv\xi$,$\ $Eq.
(\ref{21}) yields
\begin{equation}
W\left(  \gamma,\gamma^{\ast}\right)  =\frac{\left[  \left(  2\bar{n}%
_{t}+1\right)  \bar{n}_{t}-1\right]  ^{m}}{\pi\left(  2\bar{n}_{t}+1\right)
^{m+1}\left(  \bar{n}_{t}+1\right)  ^{m}}e^{-\frac{2\left\vert \gamma
\right\vert ^{2}}{2\bar{n}_{t}+1}}L_{m}\left(  -\xi\right)  .\label{24}%
\end{equation}
At the zero-temperature limit, $T\rightarrow0,\bar{n}_{t}\rightarrow0,$
Eq.(\ref{23}) reduces now into $\frac{1}{\pi}\exp(-2\left\vert \alpha
-\gamma\right\vert ^{2}),$ i.e., the WF of coherent state (a Guassian form),
which can be seen from Eq.(\ref{2}) yielding $\rho=\left\vert \alpha
\right\rangle \left\langle \alpha\right\vert $ under the condition $m=0$;
while Eq.(\ref{24}) becomes $\frac{1}{\pi}(-1)^{m}e^{-2\left\vert
\gamma\right\vert ^{2}}L_{m}(4\left\vert \gamma\right\vert ^{2}),$
corresponding to the WF of number state.

Using Eq.(\ref{21}), the WFs of the SECST are depicted in Fig.4 in phase space
with $\alpha=0.2+0.2i$ for several different values $m$ and $\bar{n}_{t}.$ It
is easy to see that the negative region of WF gradually disappears as $m$ and
$\bar{n}_{t}$ and increases.

\subsection{The marginal distributions of the SECST}

We now find the probability distribution of position or momentum%
$\vert$%
-----the marginal distributions, by performing the WF either over the variable
$y$ or the variable $x$, respectively. Using Eqs.(\ref{21}) and (\ref{22}) we
can derive (denote $\gamma=x+iy,$ $\alpha=q+ip$)%
\begin{align}
\mathrm{P}\left(  x,\bar{n}_{t}\right)   &  \equiv\int W\left(  x,y\right)
dy\nonumber\\
&  =\frac{\left(  \lambda_{t}^{2}A_{1}^{2}\right)  ^{m}C_{\alpha,m}}%
{\sqrt{2\pi\left(  2\bar{n}_{t}+1\right)  }}\frac{\left[  m!\right]
^{2}e^{-\frac{2\left(  q-x\right)  ^{2}}{2\bar{n}_{t}+1}}}{\left(  2\bar
{n}_{t}-1\right)  ^{m}}\nonumber\\
&  \times\sum_{k=0}^{m}\frac{2^{2k-m}\bar{n}_{t}^{k}}{k!\left[  (m-k)!\right]
^{2}}\left\vert H_{m-k}\left(  E_{1}\right)  \right\vert ^{2},\label{25}%
\end{align}
where $H_{m}\left(  x\right)  $ is single variable Hermite polynomial and
$E_{1}=\left[  \left(  2\bar{n}_{t}-1\right)  \alpha+2x+2ip\right]
/\sqrt{2\left(  2\bar{n}_{t}+1\right)  }$. Eq.(\ref{25}) is the marginal
distribution of WF of the SECST in \textquotedblleft$x$%
-direction\textquotedblright.

On the other hand, performing the integration over $dx$ yields the
other
marginal distribution in \textquotedblleft$y$-direction\textquotedblright,%
\begin{align}
\mathrm{P}\left(  y,\bar{n}_{t}\right)   &  =\frac{\left(  \lambda_{t}%
^{2}A_{1}^{2}\right)  ^{m}C_{\alpha,m}}{\sqrt{2\pi\left(  2\bar{n}%
_{t}+1\right)  }}\frac{\left[  m!\right]  ^{2}e^{-\frac{2(p-y)^{2}}{2\bar
{n}_{t}+1}}}{\left(  2\bar{n}_{t}-1\right)  ^{m}}\nonumber\\
&  \times\sum_{k=0}^{m}\frac{2^{2k-m}\bar{n}_{t}^{k}}{k!\left[  (m-k)!\right]
^{2}}\left\vert H_{m-k}\left(  E_{2}\right)  \right\vert ^{2},\label{e26}%
\end{align}
where $E_{2}=i\left(  2\bar{n}_{t}\alpha-\alpha+2q+2iy\right)  /[\sqrt
{2\left(  2\bar{n}_{t}+1\right)  }]$. As expected, the two marginal
distributions are both real.
\bigskip
\section{Conclusions}

In summary, we have investigated the photon statistics properties of the
SECST, described by the density matrix $\rho$ (\ref{2}). We have calculated
the matrix elements $\left\langle N\right\vert \rho\left\vert M\right\rangle $
in Fock space and the Mandel $Q$ parameter. It is found that the SECST field
exhibits a significant amount of super-PPS due to the presence of thermal
noise ($\bar{n}_{t}$) for excitation photon number $m=0$ and that, for
$m\neq0$ and a given $\left\vert \alpha\right\vert ,$ the SECST field presents
the sub-PPS when $\bar{n}_{t}$ is less than a threshold value. In addition,
the threshold value increases as $m$\ increases. We have presented the maximum
information (channel capacity) transmitted by the SECST beam. It is shown that
the maximum information transmitted increases as $m$ increases. This implies
that among the coherent signals, the eigen-number signals and the ECS in
thermal light, the last one can transmit the most information. Further, as one
of the photon statistical properties, the Wigner function and the marginal
distributions of the SECST have also been derived, from which one can clearly
see the nonclassicality. The negative region has no chance to be present when
the average photon number $\bar{n}_{t}$ of thermal noise exceeds $1/2.$ The
marginal distributions are related to the Hermite polynomial.

\begin{acknowledgments}
This work is supported by the National Natural Science Foundation of China
(Grant No 10775097). L.-Y. Hu's email address is hlyun@sjtu.edu.cn or hlyun2008@126.com.
\end{acknowledgments}

\end{document}